\documentclass[]{spie}  
\usepackage[]{graphicx}

\usepackage{graphicx}
\usepackage[space]{grffile}
\usepackage{latexsym}
\usepackage{amsfonts,amsmath,amssymb}
\usepackage{url}
\usepackage[utf8]{inputenc}
\usepackage{fancyref}
\usepackage{hyperref}
\hypersetup{colorlinks=false,pdfborder={0 0 0},}

\def\bicep{{\sc Bicep}}
\def\buno{{\sc Bicep1}}
\def\bdos{{\sc Bicep2}}
\def\btres{{\sc Bicep3}}
\def\keck{{\it Keck Array}}

\title{BICEP3: a 95\,GHz refracting telescope for degree-scale CMB polarization} 

\author{Z.~Ahmed\supit{$\dagger$a,b}, M.~Amiri\supit{c}, S.~J.~Benton\supit{d}, J.~J.~Bock\supit{e,f}, R.~Bowens-Rubin\supit{g}, I.~Buder\supit{g}, E.~Bullock\supit{h}, J.~Connors\supit{g}, J.~P.~Filippini\supit{e}, J.~A.~Grayson\supit{a,b}, M.~Halpern\supit{c}, G.~C.~Hilton\supit{i}, V.~V.~Hristov\supit{e}, H.~Hui\supit{e}, K.~D.~Irwin\supit{a,b}, J.~Kang\supit{a,b}, K.~S.~Karkare\supit{g}, E.~Karpel\supit{a}, J.~M.~Kovac\supit{g}, C.~L.~Kuo\supit{a,b}, C.~B.~Netterfield\supit{d}, H.~T.~Nguyen\supit{f}, R.~O'Brient\supit{e}, R.~W.~Ogburn~IV\supit{a,b}, C.~Pryke\supit{h}
C.~D.~Reintsema\supit{i}, S.~Richter\supit{g}, K.~L.~Thompson\supit{a,b}, A.~D.~Turner\supit{f}, A.~G.~Vieregg\supit{j}, W.~L.~K.~Wu\supit{a,b}, K.~W.~Yoon\supit{a,b}
\skiplinehalf
\supit{a}Department of Physics, Stanford University, Stanford, CA 94305, USA \\
\supit{b}Kavli Institute for Particle Astrophysics and Cosmology, SLAC National Accelerator Laboratory, Menlo Park, CA 94025, USA\\
\supit{c}Department of Physics and Astronomy, University of British Columbia, Vancouver, BC, Canada\\
\supit{d}Department of Physics, University of Toronto, Toronto, ON, Canada\\
\supit{e}Division of Physics, Mathematics, \& Astronomy, California Institute of Technology, Pasadena, CA 91125, USA\\
\supit{f}Jet Propulsion Laboratory, Pasadena, CA 91109, USA\\
\supit{g}Harvard-Smithsonian Center for Astrophysics, Cambridge, MA 02138, USA\\
\supit{h}School of Physics \& Astronomy, University of Minnesota, Minneapolis, MN 55455, USA\\
\supit{i}National Institute of Standards and Technology, Boulder, CO 80305, USA\\
\supit{j}Department of Physics, University of Chicago, Chicago, IL 60637, USA\\
}

\authorinfo{\supit{$\dagger$} Corresponding Author: Z.~Ahmed, 382 Via Pueblo Mall, Stanford, CA 94305. Email: zeesh at stanford dot edu}

\begin{document} 
\maketitle 

\begin{abstract}
\btres\ is a 550\,mm-aperture refracting telescope for polarimetry of 
radiation in the cosmic microwave background at 95\,GHz. It adopts the methodology
of \buno, \bdos\ and the \keck\ experiments --- it possesses
sufficient resolution to search for signatures of the inflation-induced
cosmic gravitational-wave background while utilizing a compact design
for ease of construction and to facilitate the characterization and
mitigation of systematics. However, \btres\ represents a significant
breakthrough in per-receiver sensitivity, with a focal plane area 5$\times$
larger than a \bdos/\keck\ receiver and faster optics ($f/1.6$ vs.
$f/2.4$). Large-aperture infrared-reflective metal-mesh filters and
infrared-absorptive cold alumina filters and lenses were developed and
implemented for its optics. The camera consists of 1280
dual-polarization pixels; each is a pair of orthogonal antenna arrays
coupled to transition-edge sensor bolometers and read out by multiplexed
SQUIDs. Upon deployment at the South Pole during the 2014-15 season,
\btres\ will have survey speed comparable to \keck\ 150\,GHz (2013), and will
significantly enhance spectral separation of primordial B-mode power
from that of possible galactic dust contamination in the \bdos\
observation patch.
\end{abstract}



\section{Introduction}
Standard $\Lambda$CDM cosmology provides a successful and self-consistent framework that explains experimental data sampling various cosmic epochs -- light element abundances match those predicted by Big Bang Nucleosynthesis\cite{fields_big_2006}, relic radiation from the last photon-electron scattering is observed as the Cosmic Microwave Background (CMB) matching theory\cite{hinshaw_nine-year_2013, planck_collaboration_planck_2013}, observations of large-scale structure in the later history of the Universe appear to arise from the expected primordial power spectrum\cite{eisenstein_detection_2005}. However, standard $\Lambda$CDM cosmology fails to explain the  homogeneity and isotropy of the observable Universe, as well as the observed geometric flatness of the Universe. Inflation, an exponential expansion of space-time at the earliest epochs, resolves these issues and  provides a mechanism to stretch primordial quantum fluctuations in the early Universe to seed structure we observe today\cite{starobinsky_new_1980, guth_inflationary_1981, linde_new_1982}. The paradigm of Inflation already enjoys circumstantial evidence, most prominently by the observation of a power law spectrum of perturbations and a measured departure from exact scale-invariance of perturbations\cite{planck_collaboration_planck_2013}. However, inflation models also generically predict a stochastic gravitational wave background generated by magnification of quantum perturbations of the gravitational field\cite{starobinsky_spectrum_1979, rubakov_graviton_1982, fabbri_effect_1983}. The ratio of the amplitude of the tensor perturbations that generate primordial gravitational waves to the amplitude of scalar perturbations that seed  structure is called `$r$', and is used to characterize inflationary models as well as the energy scale of inflation.  Primordial gravitational waves are expected to interact with the cosmic microwave background at the surface of last scattering and imprint a parity-odd linear polarization at degree angular scales\cite{seljak_signature_1997, kamionkowski_probe_1997}. A detection of this primordial  `B-mode' polarization would provide direct evidence for inflation and gravitational waves, and would provide the first direct hints of the quantum nature of gravity.  

However, B-mode polarization is also generated by other mechanisms. Lensing of the parity-even or `E-mode' component of CMB polarization by large-scale structure\cite{zaldarriaga_gravitational_1998} is one such source. This component is quantifiable and can be subtracted from spectra and maps\cite{seljak_gravitational_2004}. A second source of contamination of B-modes is  dust in our own galaxy, along the line of sight to the CMB. Removal of this contaminant requires multi-color observations to enable spectral discrimination between CMB signal and galactic foregrounds\cite{dunkley_prospects_2009}. The level of challenge presented by these contaminants to the measurement of primordial B-mode signal depends on the amplitude of the contaminants relative to the amplitude of the primordial signal.

Recently, \bdos\ made the first detection of B-mode power on degree angular scales at 150 GHz\cite{bicep2_collaboration_detection_2014}. The data fit well a model that includes $\Lambda$CDM, expected B-mode power from lensing and $r=0.2$, suggesting a detection of primordial gravitational waves generated by inflation. However, a lack of sufficiently constraining spectral data from the \bicep/{\it Keck} collaboration or other publicly available datasets permits the high value of $r$ suggested by the fit to be lowered by an uncertain amount to accommodate galactic dust contamination.  The path forward to deconvolving the foregrounds from the primordial signal lies in observations of the CMB at multiple frequencies. Several existing and planned experiments are on track to do this in the next few years.  Two receivers of the \keck\ have begun re-observing the \bdos\ sky patch at 95 GHz\cite{buder_bicep2_2014}. Also, the Planck collaboration intends to release data products and analysis results from its multi-frequency polarization data within the year. The data from these experiments might have the sensitivity required to resolve the foreground uncertainties in the \bdos\ results.

In parallel, the \bicep/{\it Keck}  collaboration has continued R\&D towards even more sensitive instruments to cross check the \bdos\ results and measure CMB B-mode power at degree scales at higher significance. These efforts have resulted in a new 95\,GHz instrument called \btres. \btres\ presents a breakthrough in CMB polarimetry throughput and sensitivity for refracting telescopes. This single instrument doubles the traditional \bdos/\keck\ aperture and combines the detector count of five \bdos-like receivers or the entire \keck. This has been achieved by implementing modular detector-array packaging, improved infrared filtering for large clear aperture, and fast alumina optics with a novel implementation of anti-reflection coating. These proceedings discuss the design of \btres.

\section{Detectors, Readout \& Camera} 
\begin{figure}[tbp]
\centering
\includegraphics[width=0.45\textwidth]{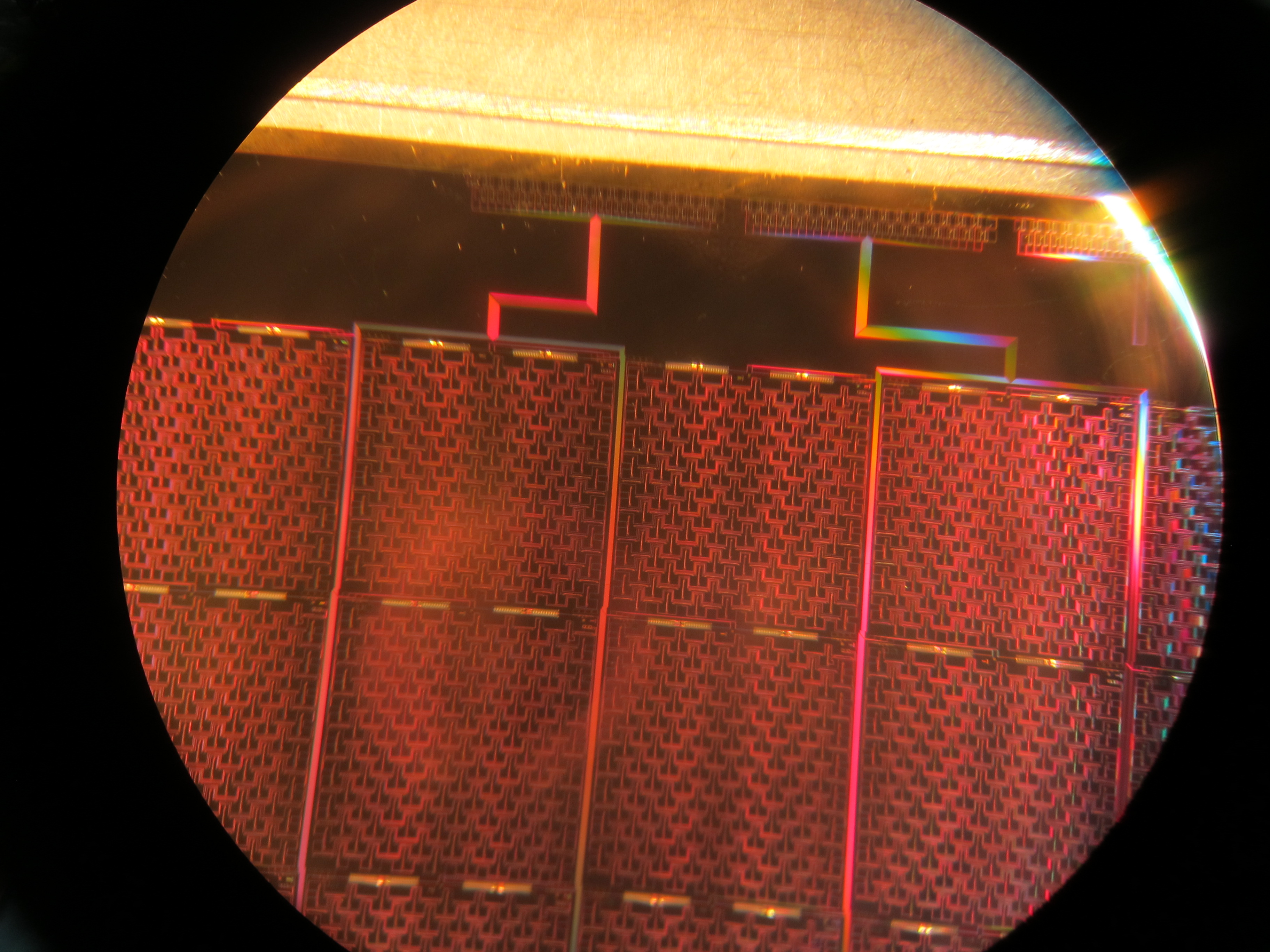}
\includegraphics[width=0.45\textwidth]{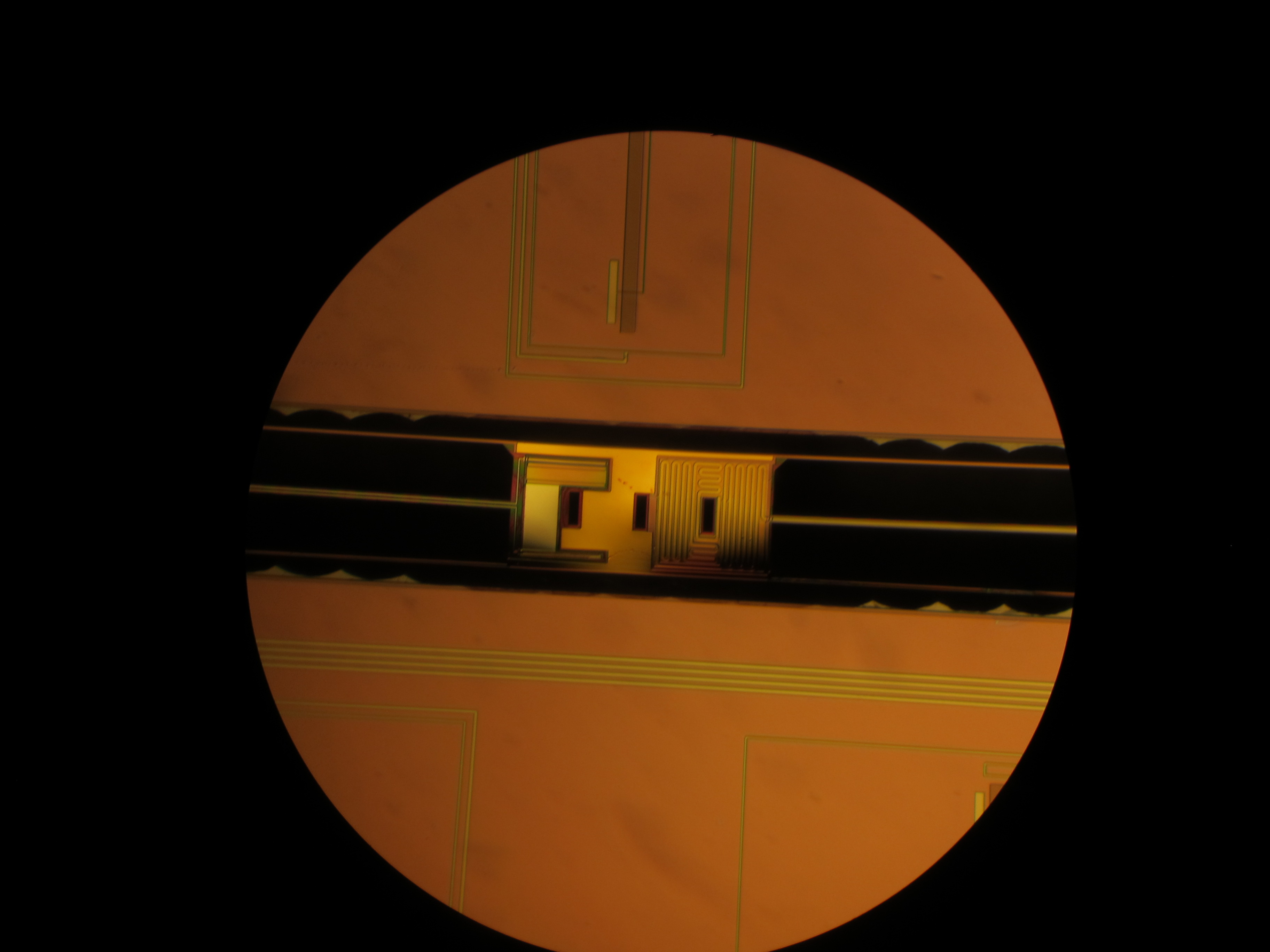}
\caption{{\it Left}: An array of \btres\ dual-polarization pixels under a microscope. Each square unit is a pixel and consists of two 12x12 network of slot antennas, one each for an orthogonal polarization. {\it Right}: A transition edge sensor (TES) island under a microscope. Each dual-polarization pixel has two TES islands, one for each slot-antenna network. Power is introduced through the long microwave strip entering on the right side of the island and absorbed in an Au absorber (meander). The incoming power changes the resistance of the Ti TES on the left side of the island.}
\label{fig:pixel}
\end{figure}

\btres\ uses polarization-sensitive millimeter-wave detector technology developed at JPL/Caltech, and also used in \bdos/\keck\cite{kuo_antenna-coupled_2008, orlando_antenna-coupled_2010}. Each camera pixel combines two orthogonal slot-antenna networks, band-defining filters, absorbers, and transition-edge sensor (TES) bolometers on a single silicon substrate, as shown in Figure \ref{fig:pixel}. Single-moded, diffraction-limited beams are obtained without the need for feed horns by in-phase summation of power collected by antennae in a corporate feed network. The feed network is designed to minimize beam pointing differences between the two polarizations for a pixel and uses  gaussian-tapered illumination to suppress side lobes \cite{obrient_antenna-coupled_2012}. The microwave power collected in a single polarization feed network is deposited on a gold absorber coupled to a TES bolometer. To control the bolometer conductance to the thermal bath, the absorber and bolometer sit separately on a silicon nitride island suspended from the substrate. The silicon tiles are held at $\sim280$\,mK and the titanium TESs self-heat to their superconducting transition temperature in the range 470--530\,mK. Fluctuations in incoming power alter the TES resistance, which in turn changes the current through the bias circuit. An inductor in series with the TES couples the changes in current to superconducting quantum interference devices (SQUIDs). \btres's SQUIDs, designated MUX11d, were developed and fabricated at NIST, Boulder\cite{irwin_time-division_2002}. The SQUIDs are arranged in time-domain multiplexing arrays at cold temperatures. Room temperature electronic complete a fast negative-feedback loop to cancel magnetic flux changes in the series inductors. The room temperature electronics, called Multichannel Electronics (MCE)\cite{battistelli_functional_2008} were developed at the University of British Columbia and are implemented in several other experiments.

\begin{figure}[tbp]
\centering
\includegraphics[width=0.45\textwidth]{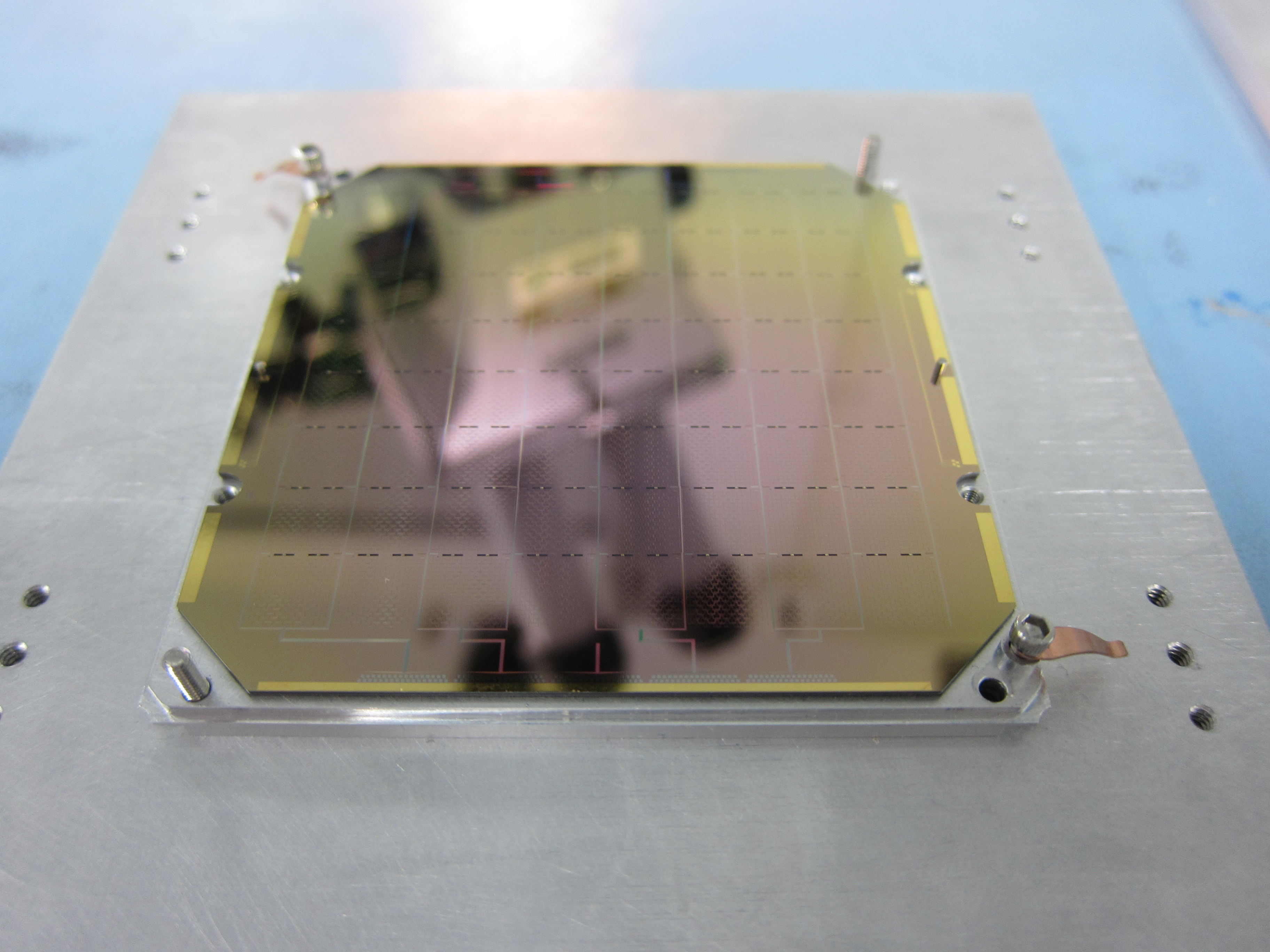}
\includegraphics[width=0.45\textwidth]{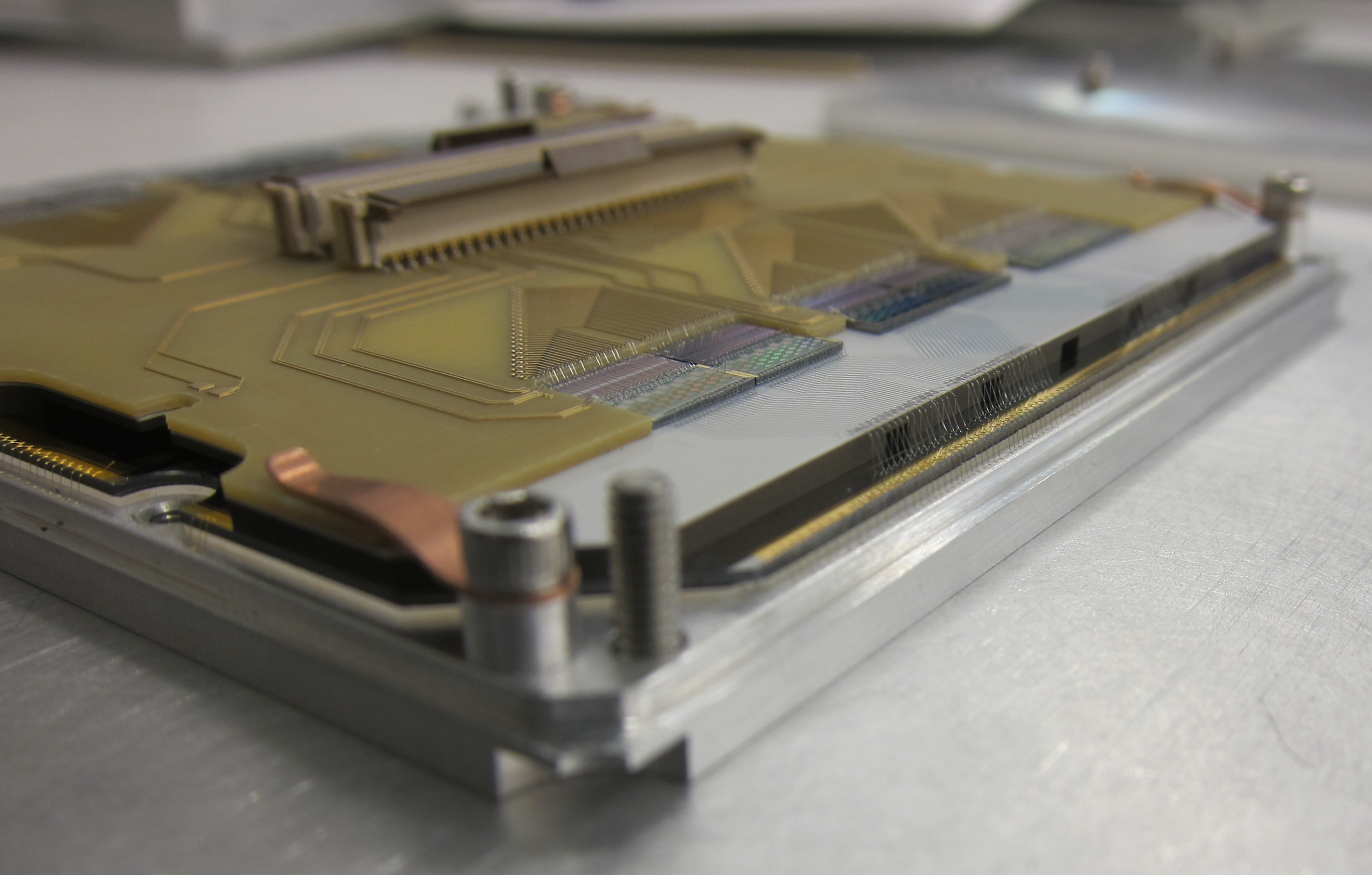}
\includegraphics[width=0.6\textwidth]{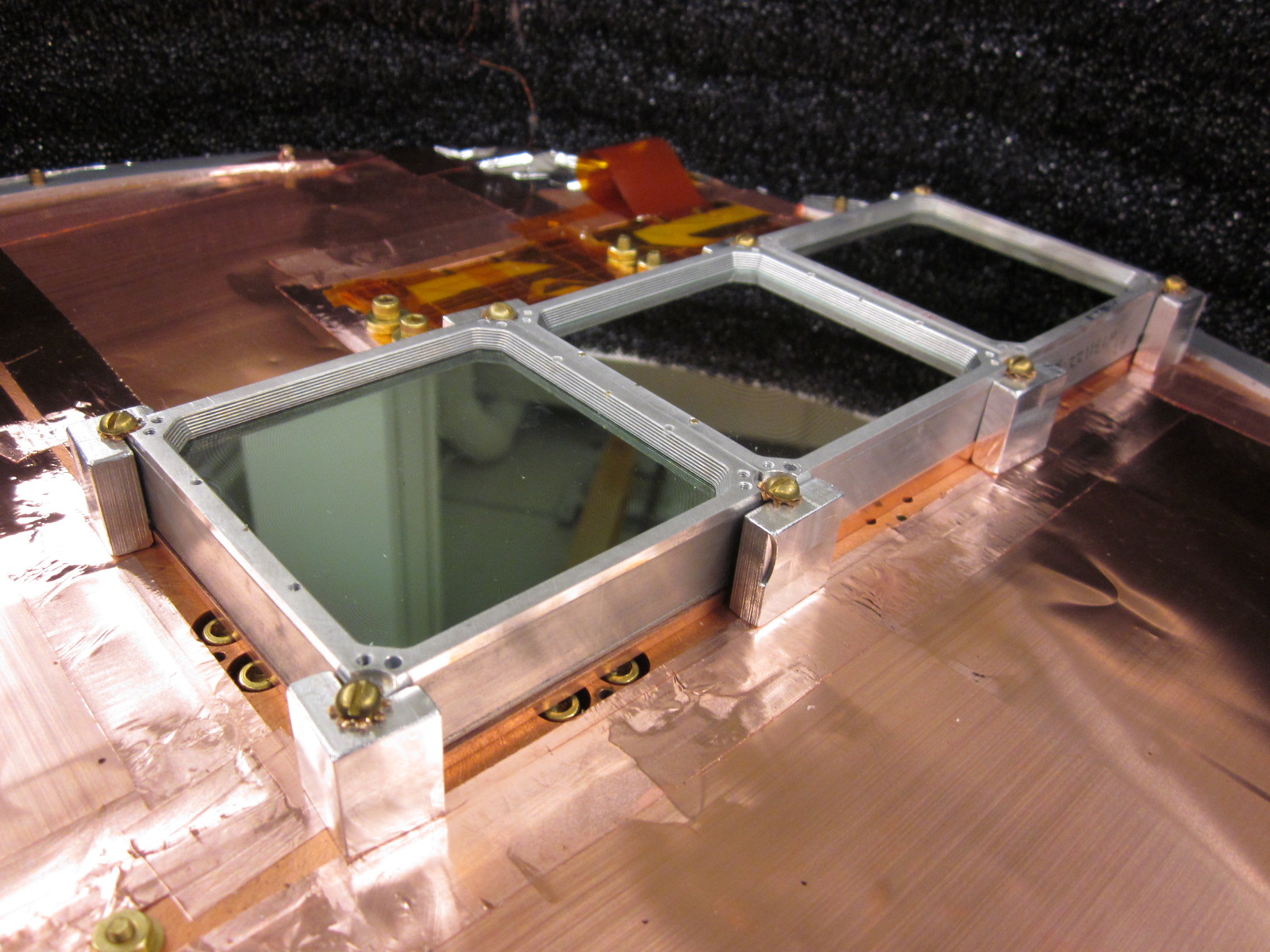}
\caption{{\it Top Left}: A \btres\ detector-array tile undergoing assembly into a module. {\it Top Right}: A \btres\ module without its niobium enclosure. The detectors face downward in this picture, exposing the circuit boards, SQUID multiplexing chips, and wire bonds. A standard connector on top of the PCB provides electrical connections to the outside. {\it Bottom}: Three modules installed in the \btres\ receiver for testing.}
\label{fig:module}
\end{figure}


Depending on the desired frequency, $\sim$30--150 dual-polarization pixels can be patterned onto single 3"$\times$ 3" Si tiles via photolithography. Such detector-array tiles can be mass produced. \btres's focal plane can accommodate 20 detector-array tiles, packaged into `plug-and-play' modules as shown in Figure \ref{fig:module}. Each 95\,GHz module for \btres\ has 64 dual-polarization pixels (128 bolometers) for a total of 1280 pixels (2560 bolometers). A module consists of a detector-array tile and its supporting cold electronics, such that hand-affixed cryogenic flex ribbon cable and mechanical fasteners are the only required connections between the module and the focal plane heat sink. The module housings are fabricated using niobium to provide a superconducting magnetic shield around the SQUIDs at  operating temperatures. The detector-array modules present a significant simplification from the \bdos/\keck\ focal plane design; detector-array tiles of those experiments were directly wire bonded to the focal plane and were challenging to modify or repair --- a scheme nearly unviable for a focal plane with 20 detector-array tiles.



\section{Receiver Design} 

\begin{figure}[tbp]
\centering
\includegraphics[width=0.8\textwidth]{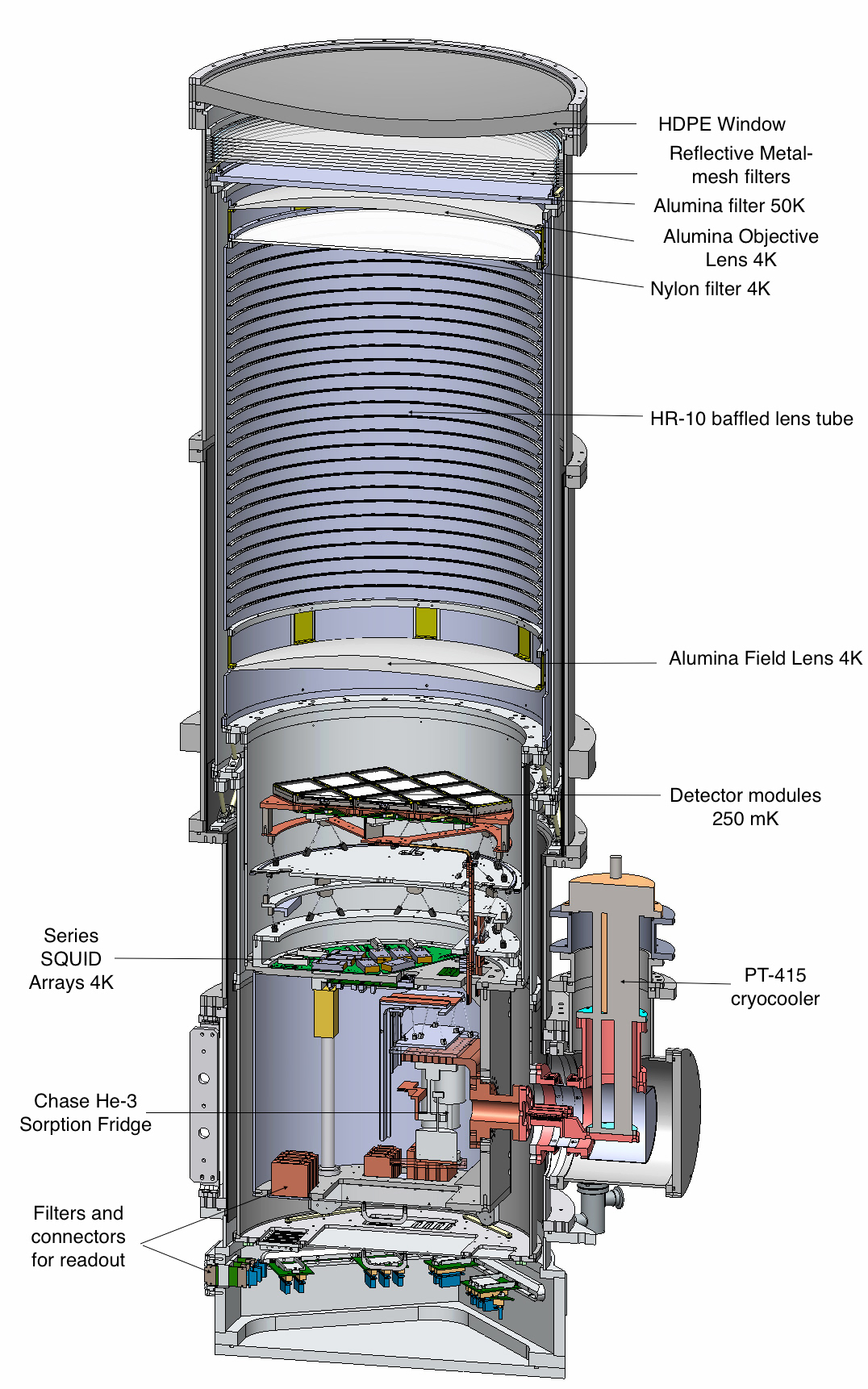}
\caption{Cross section of \btres\ receiver showing various components.}
\label{fig:receiver}
\end{figure}

The \btres\ receiver is housed in a custom-designed vacuum cryostat 2.4\,m tall and 0.73\,m in diameter. It has been designed to fit into the BICEP mount at the South Pole. In addition to azimuth and elevation rotation the mount has rotation about its optical axis for instrumental systematics control. The \btres\ receiver design is modeled on \buno\cite{yoon_robinson_2006} and \bdos\cite{bicep2_collaboration_bicep2_2014}. A CAD cross-section is displayed in Figure \ref{fig:receiver}. The receiver was designed and constructed at Stanford University and SLAC. The vacuum jacket is sectioned into three lengthwise segments for ease of access and assembly/disassembly, enabling rapid test and operation cycles. The top section houses an HDPE vacuum window, a stack of reflective metal-mesh infrared shaders, an infrared-absorptive alumina filter at 50\,K, alumina lenses at 4\,K, and a nylon filter also at 4\,K. The middle section houses the sub-kelvin focal plane with the associated thermo-mechanical structure. The lower section houses the cryogenic system and cold electronics to support the focal plane, as well as ports to warm electronics. Inside the vacuum jacket, the cryostat is radially partitioned by  4\,K and 50\,K radiation shields. The  4\,K and 50\,K volumes are nested using G-10 trusses in the middle lengthwise section and G-10 tensile straps on top and bottom lengthwise sections, providing structural support and constraints while maintaining low thermal conductivity between temperature stages. The 4\,K to sub-K structures supporting the focal plane consist of carbon fiber trusses for its high ratio of stiffness to thermal conductivity  below 4\,K. 

\section{Thermal Architecture} 
\btres's thermal architecture minimizes non-CMB loading on the detectors, just as its predecessors. The optics, detectors and supporting cold electronics are housed in nested 50\,K and 4\,K cold-temperature shields, heat sunk respectively to the first and second stages of a Cryomech PT-415 cooler. The cooling capacity of the cryocooler at these stages is 40\,W and 1.5\,W respectively.  This would have been a crippling limitation in advancing from \bdos/\keck's  aperture size to \btres's. In particular, a $\sim$60-cm window presents $\sim$165\,W of infrared loading, peaked at $\sim$10\,$\mu$m wavelength from room temperature loading. 

\btres\ uses custom-developed reflective metal-mesh infrared shaders\cite{ahmed_large-area_2014} to reduce sky infrared loading to an acceptable level for the subsequent absorptive filters. The filters material is 610\,mm-diameter, 3.5\,$\mu$m-thick mylar coated with thin-film aluminum. The aluminum is ablated by a 355\,nm UV laser to leave a capacitive grid of squares for low-pass filtering. For \btres, we use a stack of 8 such filters --  two with 40\,$\mu$m-side squares and 55\,$\mu$m pitch, four with 50\,$\mu$m-side squares and 80\,$\mu$m pitch, and two with 90\,$\mu$m-side squares and 150\,$\mu$m pitch. These filters have staggered cut-off frequencies close to $\sim$1\,THz and reduce the power incident on the 50\,K stage to $\sim$15\,W.   

The residual power is absorbed by a 1\,cm-thick high-purity, low-loss alumina disc at 50\,K and a 5\,mm-thick cast Nylon disc at 4\,K. Both have diameter greater than 550\,mm. Alumina is favored over PTFE filters used in \bdos/\keck\ because alumina provides O(100) times better thermal conductivity when heat sunk at 50\,K and has a low-pass frequency cut-off of $\sim$1\,THz\cite{inoue_cryogenic_2013}. All \btres\ alumina is sourced from a single batch of 99.6\% pure stock (AD-996 SI) manufactured by CoorsTek Inc\footnote{http://www.coorstek.com/materials/ceramics/alumina.php}.  We made extensive index and absorption tests of alumina samples from several manufacturers before picking this batch of alumina for \btres.

Sub-kelvin cooling for the detectors is provided by a 33-L three-stage helium sorption fridge from Chase Research Cryogenics at 2\,K (heat exchanger), 350\,mK (intercooler) and 250\,mK (ultra cooler). Detector modules are mounted on a copper focal plane heat sunk to the ultra cooler via a flexible high-purity copper-foil heat strap and stainless steel blocks.  The intercooler and heat exchanger provide buffer points to intercept heat load from 4K to the focal plane.


Thermal monitoring for the cryostat is done using calibrated diodes and resistance thermometers at the radiation shields, critical cryogenic junctions, and at the edges of all optical elements. The focal plane temperature is maintained at a stable temperature by passive and active filtering similar to \bdos\cite{bicep2_collaboration_bicep2_2014}. The stainless steel blocks positioned in the thermal path between the ultra cooler and the focal plane act as a passive low-pass filter. Active control is implemented in a feedback loop using a neutron-transmutation doped (NTD) Ge thermometer and a resistive heater each on the two sides of the stainless steel filter. The ADC/DAC for thermal monitoring and control was designed at the California Institute of Technology and Stanford University. It was built at the University of Minnesota, incorporating the BLASTBus2 data acquisition system\cite{benton_blastbus_2014} from the University of Toronto. 

\section{Optics} 
Similar to all \bicep/{\it Keck} experiments, \btres's  optical elements are housed inside the receiver cryostat and cooled to $\sim4$\,K to minimize loading on the detectors. The telescope is a simple two-lens, diffraction-limited, on-axis telecentric refractor to keep aberration and distortions subdominant.  It differs from its predecessors in that its clear aperture is twice as large and its detectors and optics have been sized and situated to enable faster images and greater optical throughput. This results in f-ratio $f/1.6$ compared to $f/2.4$ for \bdos/\keck. The lenses are 580\,mm in   diameter and are made of the same type of alumina as the absorptive filter at 50\,K. The high refractive index of alumina ($n=3.1$ at $\lambda=3.16\,$mm) allows the lenses to be thinner than equivalent lenses of high-density polyethylene (HDPE). Alumina lenses are also more dimensionally stable than HDPE in fabrication and operation. 

Inactive areas of the 4\,K optics tube are covered in epoxy-encapsulated Eccosorb HR-10 microwave absorber. The HR-10 is cut and patterned into ridges to break millimeter-wavelength specular reflection and limit shallow incidence angle reflections that could illuminate side lobes to sub-percent amplitude. Epoxy-encapsulated HR-10 is also used to define the optical stop for the system behind the objective lens. 

\begin{figure}[tbp]
\centering
\includegraphics[width=0.35\textwidth]{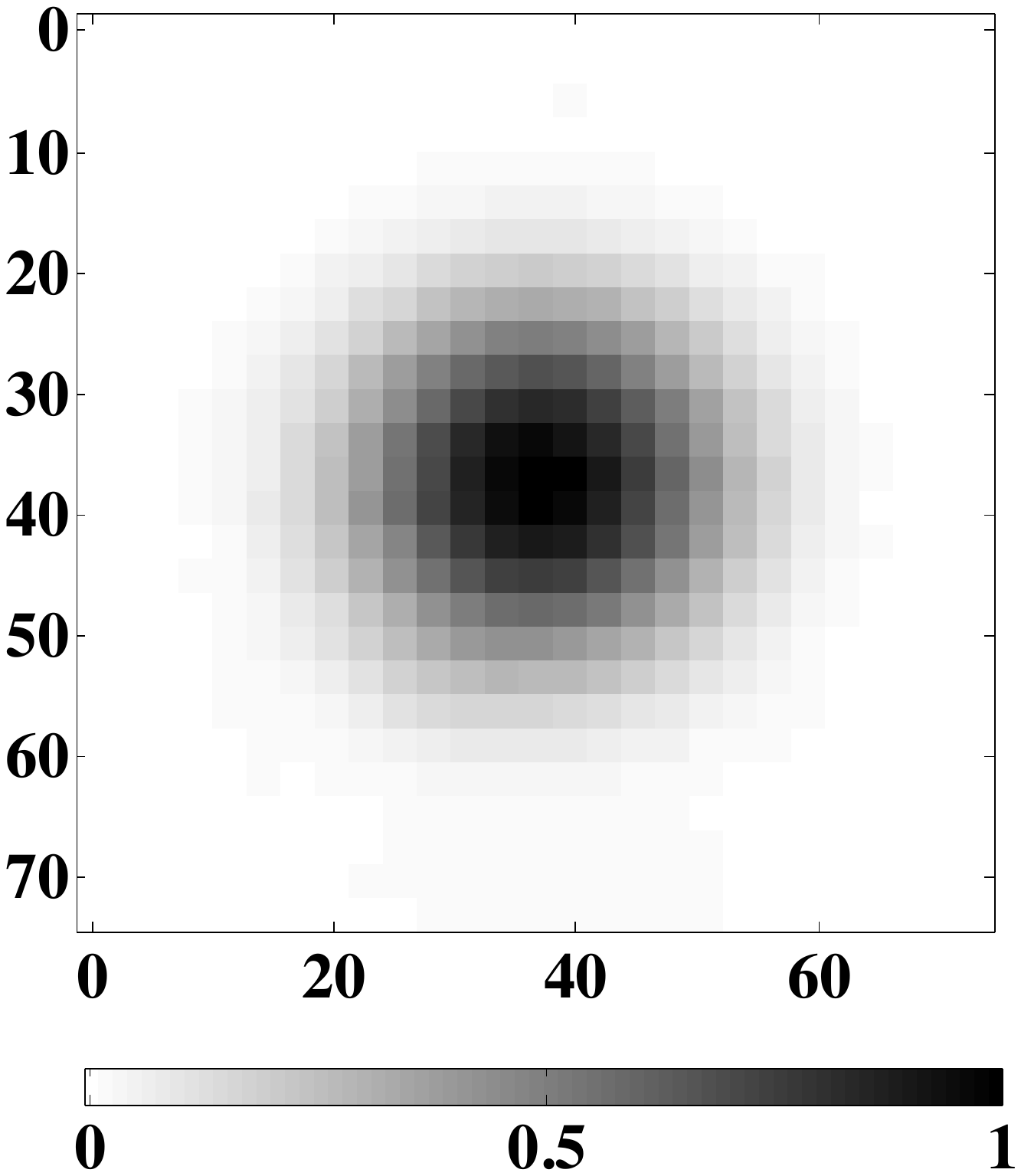}
\includegraphics[width=0.55\textwidth]{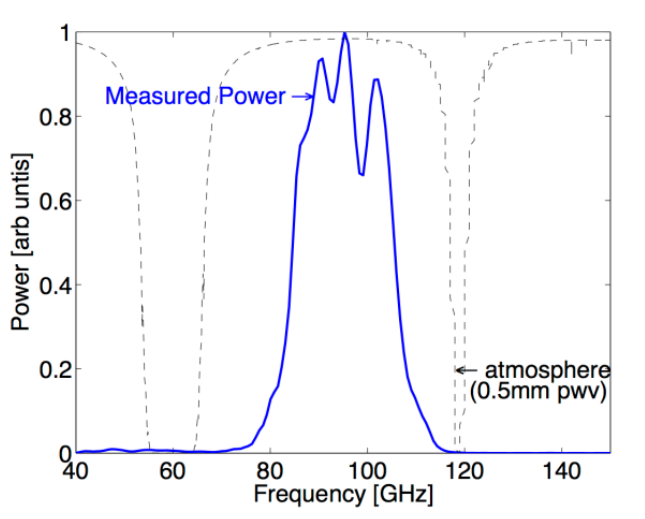}
\caption{Left: Normalized beam pattern of a typical \btres\ detector as measured by a refocusing to quasi-far field through a warm lens. Horizontal and vertical axes are in units of arc minutes. Right: Bandpass spectrum for a typical \btres\ detector as measured by a Martin-Puplett Fourier Transform Spectrometer (FTS). Also shown for reference is the atmospheric transmission function at the South Pole.}
\label{fig:qffbm}
\end{figure}

We have measured the beam characteristics of \btres\ detectors through this optical chain in quasi-far field mode by using a mirror and a collimating lens to refocus the near-field beam to a scanning thermal source. The detectors see a modulated signal between room temperature and a few hundred degrees above that using a chopper wheel.  The FWHM of a few beams has been measured to be $<\sim25$ arcmin at 95\,GHz. A beam map is shown in Figure \ref{fig:qffbm}. The beams appear well behaved and beam concentricity of orthogonal polarizations was measured to be $\sim$2\% of the FWHM.   Additionally, we  characterized the bandpass filters on the pixel antenna-summing trees using a Martin-Puplett Fourier Transform Spectrometer (FTS)\cite{karkare_keck_2014}. A typical bandpass is shown in Figure \ref{fig:qffbm}.

\section{Anti-reflection coating for Alumina Optics}
The large aperture alumina lenses have high index, and thus can have high reflectance. Without a suitable anti-reflection coating (ARC), the \btres\ optical chain would have a total reflectance of up to $\sim85\%$. Rosen et al. have demonstrated the viability of Stycast epoxies as ARC for small optical elements of sizes $\sim10$\,mm \cite{rosen_epoxy-based_2013}.  For larger coated areas, we observed the ARC to crack and de-adhere because of differential thermal expansion between it and the alumina. For \btres\, we developed a custom epoxy mixture to provide the correct refractive index for maximum transmission, and also developed dicing techniques to achieve cryogenically stable ARC for for large-area alumina lenses and filters. 

\begin{figure}[tbp]
\centering
\includegraphics[width=0.45\textwidth]{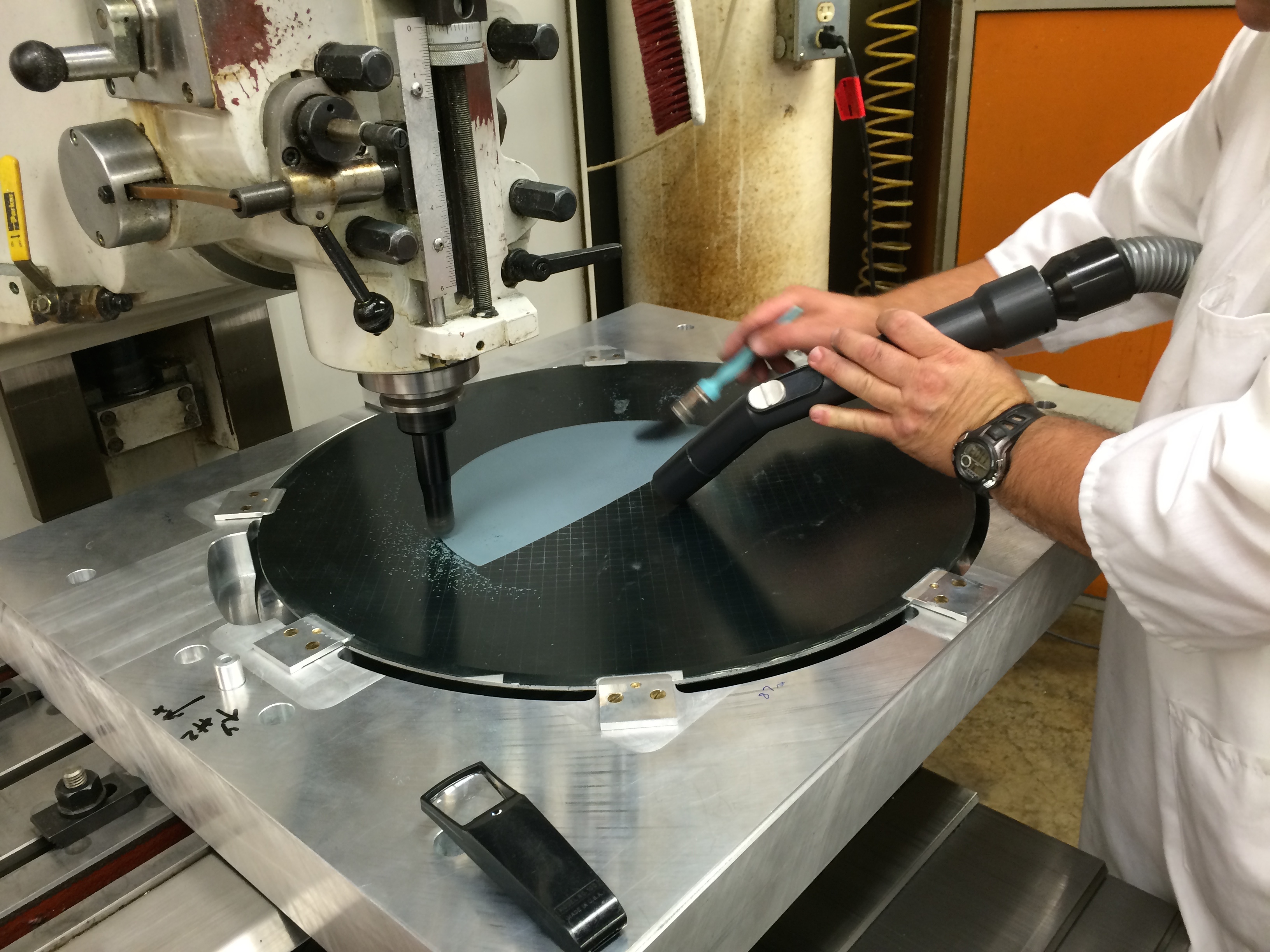}
\includegraphics[width=0.45\textwidth]{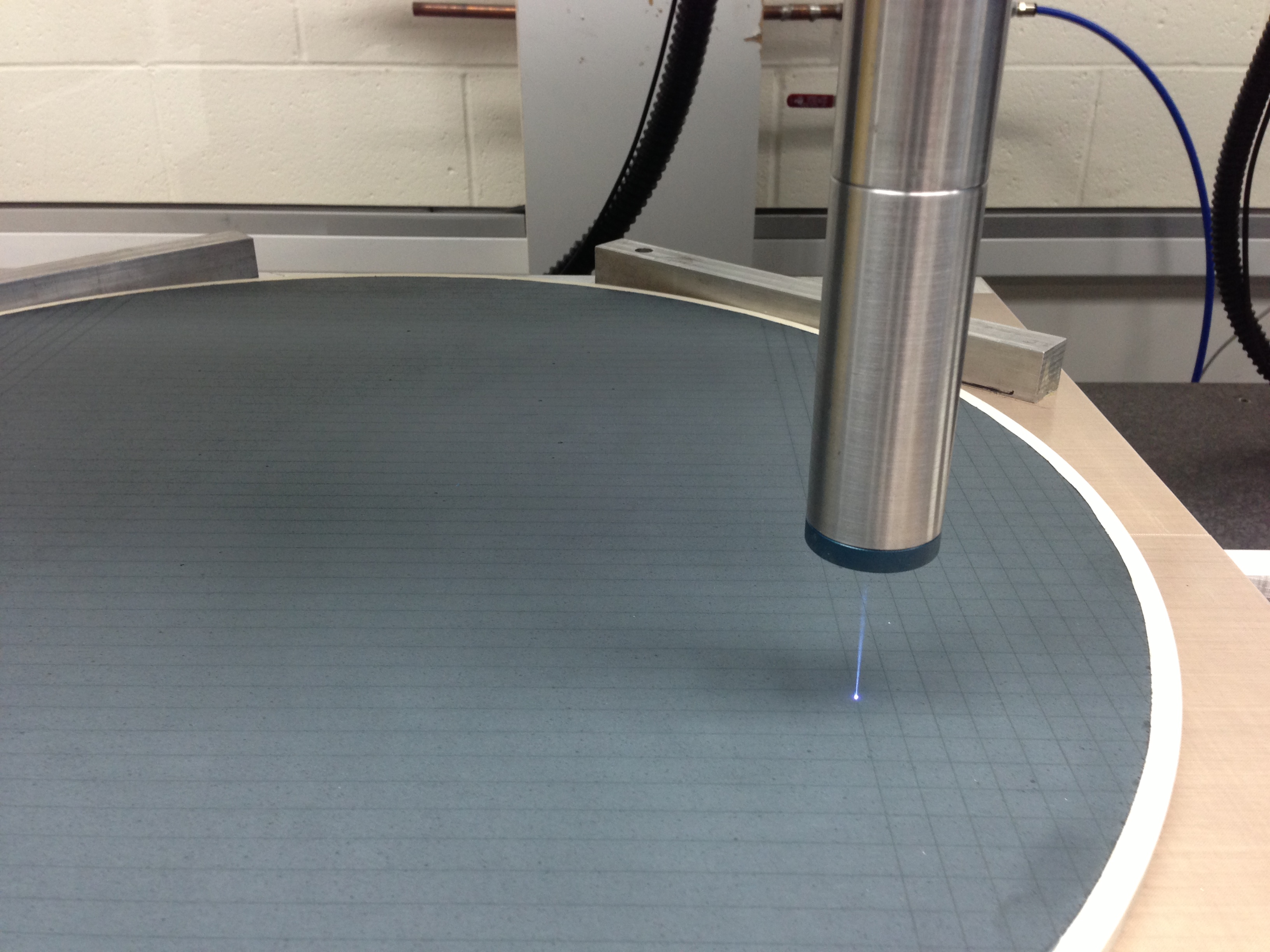}
\caption{Left: CNC machining of anti-reflection coating (ARC) on concave surface of lens to appropriate thickness for maximum transmission.  Right: Laser stress-relieving of ARC on a filter to prevent de-adhering from differential thermal contraction during cryogenic cycling.}
\label{fig:arc}
\end{figure}

As an ARC for our alumina optics we use a blend of Stycast 1090 and Stycast 2850 in a 3:2 ratio by volume, activated by Catalyst 23LV. We measured this mixture to have an index $1.74\pm0.01$ at 95\,GHz when cured. For consistency between different ARCs, the epoxy quantities are individually measured to 0.5\% uncertainty in weight and mixed mechanically for a preset time. Furthermore, the mixing is done under vacuum to facilitate outgassing and to prevent bubbles from forming inside the ARC. We have established by index measurement that mixture separation is negligible during the epoxy set up time.  The epoxy mixture is deposited on the surface to be coated, and rough molded to 1\,mm thickness using a precision aluminum mold. Once cured, the ARC must be reduced to the correct thickness. This is 0.452\,mm for maximum in-band transmission for the epoxy mixture we use on the \btres\ alumina optics. For a flat filter, the required thickness is achieved by abrasive polishing to $\pm10\,\mu$m uncertainty. For the lenses, the ARC is directly machined to $\pm25\,\mu$m uncertainty with a diamond tool on a CNC end mill as shown in Figure \ref{fig:arc}. For reference, a 25\,$\mu$m systematic offset on 4 ARC surfaces for the lenses represents a 1\% reduction in transmission for our 95\,GHz band, compared to maximum of 98.2\% for a perfect ARC. The achieved accuracy for direct machining is made possible by characterizing the lens surfaces and features with a coordinate-measuring machine (CMM) to 2\,$\mu$m-uncertainty before ARC deposition, and the establishment of a common coordinate system between the CMM and CNC to 5\,$\mu$m uncertainty in the vertical direction and $\pm25\,\mu$m in the plane perpendicular to the vertical.  Finally, the ARC is diced into 1\,cm$^2$ patches using a 355\,nm UV laser with 30\,$\mu$m spot size and depth of cut down to the alumina substrate, as shown in Figure \ref{fig:arc}. This work is performed by Laserod Technologies LLC\footnote{http://www.laserod.com}. The dicing procedure relieves mechanical stresses from differential thermal contraction at cryogenic temperatures. Laser-diced ARC prepared in this way on \btres\ alumina optical elements has remained intact under repeated cryogenic cycling.



\section{Operation \& Science}
At the time of submission of these proceedings, the \btres\ receiver
has been built and is being tested in a replica of the BICEP mount at
Harvard University.  \btres\ will be deployed to the South Pole during
the austral summer of 2014--15.  The instantaneous field of view of
\btres\ is 28$^{\circ}$, compared to $\sim $20$^{\circ}$ for \bdos.  The observing patch of
\btres\ can be additionally expanded by scanning between wider endpoints
in azimuth relative to the \bdos/\keck\ observing strategy.  By
mapping a larger field, \btres\ will measure a larger number of
multipole modes than \bdos/\keck, and can therefore obtain a
lower sample variance.  Although \btres\ observes at a larger
wavelength, its optical design gives a slightly smaller beam size of $\sim$25\,arcmin.  This results in an improvement in resolution compared to the 30\,arcmin beams of \bdos/\keck\ at 150 GHz, and a more
substantial improvement compared to the 43\,arcmin beams of the \keck\ receivers at 95\,GHz.

Although measurements of \btres\ performance on a cold CMB sky are
not yet available, we can estimate its sensitivity from the \keck\
95\,GHz detectors.   This gives a projected instantaneous sensitivity of
$\sim7\,\mu$K$\sqrt{s}$ for 1280 pixels.  After two years of observation over the
\bdos\ patch, the \btres\ map depth at 95\,GHz will be approximately equal to the 150\,GHz map depth from the \keck\ at that time.  These maps will provide the tightest spectral constraints on the contributions of CMB
and foregrounds to the observed degree-scale B-mode signal.


\section{CONCLUSION}
In these proceedings, we describe the design of \btres, a new refracting telescope for CMB polarimetry at degree angular scales for high sensitivity to inflationary B-mode power. \btres\ uses modular design principles to simplify large-area focal plane assembly and utilizes new breakthroughs in infrared filtering, high-index optics and anti-reflection coating to double aperture size and throughput. This progress has enabled \btres\ to house a camera of 1280 photon noise-limited dual-polarization pixels at 95\,GHz. \btres\ will deploy at the South Pole in the austral summer of 2014-15 and commence observations.

\section*{ACKNOWLEDGMENTS} 
This work is made possible through support from the National Science Foundation (grant nos. 1313158, 1313010, 1313062, 1313287, 1056465, 0960243), the SLAC Laboratory Directed Research and Development Fund, the Canada Foundation for Innovation, and the British Columbia Development Fund. The development of detector technology was supported by the JPL Research and Technology Development Fund and grants 06-ARPA206-0040,
10-SAT10-0017 and 12-SAT12-0031 from the NASA APRA and SAT programs. The development and testing of detector modules was supported by the Gordon and Betty Moore Foundation. 

We thank Ryne Tacker and Grey Brooks at Laserod Technologies LLC, Ed Reese and Keith Caban at SLAC Precision Measurement \& Inspection, and Mehmet Solyali and Karlheinz Merkle at the Stanford Physics Machine Shop for their cooperation, persistence and ingenuity in addressing technical challenges. We are grateful to Irene Coyle, Kathy Deniston, Donna Hernandez, and Dana Volponi for administrative support. Finally, we thanks members of the larger \bicep/\keck\ family for valuable discussions and sharing decades of experience with us.



\end{document}